\documentclass[showpacs,twocolumn,aps,preprintnumbers,letterpaper]{revtex4}
\usepackage{amsmath,amssymb}
\usepackage{epsfig}
\usepackage{graphicx}
\usepackage{amsmath}
\usepackage{slashed}
\usepackage{amsfonts}
\usepackage{epstopdf}
\usepackage{color}

\newcommand{\be}{\begin{equation}}
\newcommand{\ee}{\end{equation}}
\newcommand{\bear}{\begin{eqnarray}}
\newcommand{\eear}{\end{eqnarray}}
\newcommand{\ba}{\begin{array}}
\newcommand{\ea}{\end{array}}

\def\be{\begin{eqnarray}}
\def\ee{\end{eqnarray}}

\def\roughly#1{\mathrel{\raise.3ex\hbox{$#1$\kern-.75em%
\lower1ex\hbox{$\sim$}}}}

\begin{document}

\title{Quasi-parton distribution functions:\\
two-dimensional scalar and spinor QCD}

\author{Xiangdong Ji$^{1,2}$, Yizhuang Liu$^3$  and Ismail Zahed$^3$}

\email{xji@physics.umd.edu}
\email{yizhuang.liu@stonybrook.edu}
\email{ismail.zahed@stonybrook.edu}
\affiliation{$^1$Department of Physics and Astronomy, University of Maryland, College Park, Maryland 20742, USA\\
$^2$ Tsung-Dao Lee Institute, Shanghai Jiao University, Shanghai, 200240, China\\
$^3$Department of Physics and Astronomy, Stony Brook University, Stony Brook, New York 11794-3800, USA}



\date{\today}
\begin{abstract}
We construct the quasi-parton distributions of mesons for two-dimensional QCD with either scalar or spinor quarks
using the $1/N_c$ expansion. We show that in the infinite momentum limit, the parton distribution function
is recovered in both leading and sub-leading order in $1/N_c$.

\end{abstract}


\maketitle

\setcounter{footnote}{0}


\section{Introduction}

Light cone distribution amplitudes are central to the description of hard exclusive processes
with large momentum transfer. They account for the non-perturbative quark and gluon content 
of a hadron in the infinite momentum frame. Using factorization, hard cross sections can be 
split into soft partonic distributions convoluted with perturbativly calculable processes. The
partonic distributions are inherently non-perturbative. They are currently estimated using 
experiments, lattice simulations or models. 

Recently one of us~\cite{JI1} has suggested that the light cone hadronic wavefunctions 
can be recovered from Euclidean correlators in hadronic states using instead 
quasi-parton distribution functions  through pertinent renormalization in the infinite momentum limit. 
Preliminary lattice simulations have proven very promising~\cite{JI2,ALEX}. The purpose of this
letter is to explore this construct in two-dimensional scalar and spinor QCD in the non-perturbative $1/N_c$
expansion.

Two-dimensional scalar QCD has a smooth large $N_c$ limit with a confining spectrum~\cite{TSAO,TAM,GRIN}. In this
model the current correlators exhibits many features of four-dimensional QCD in contrast to two-dimensional spinor QCD~\cite{HOOFT}. 
In the deep inelastic  regime  the results exhibit expected scaling laws, and are overall in support of the Feynman partonic picture
and the light cone expansion. 
In this paper, these two models will be used interchangeably to test the concept of the
quasi-distributions in a non-perturbative context, as they differ by a minor change in the algebra of the pertinent bosonic operators. 
Specifically, we  construct the quasi-parton distributions for both scalar and spinor QCD in leading and subleading order in $1/N_c$ and show that they merge with the expected light cone distributions
in the infinite momentum limit without additional renormalization. Our leading conclusion for two-dimensional spinor QCD is in agreement
with a recent study~\cite{RECENT}.

The organization of the paper is as follows: in section II we discuss a canonical quantization of
two-dimensional scalar QCD in the axial gauge. We make explicit the Hamiltonian  of the model in leading
order in $1/N_c$ using bosonized fields. Some renormalization issues are also discussed. 
In section III we explicit the wavefunction for scalar QCD in the light cone limit. 
In section IV we construct the quasi-parton distribution function in leading order in $1/N_c$, and show that
it reduces to the light cone wavefunction in the infinite momentum limit. 
We also discuss the leading  correction in $1/P$.  In section V, we show how to generalize
the bosonization scheme algebraically for both scalar and spinor QCD, and use it for a systematic
organization of the operators in $1/N_c$. This scheme is used 
in section VI, VII to correct
the light cone parton distribution and quasi-distribution  in spinor two-dimensional QCD through standard
perturbation theory. We show that  the subleading  corrections to the quasi-parton distribution function merges with the parton distribution function in the infinite momentum limit without renormalization. Our conclusions are in section VIII.
In the Appendix we summarized some elements of two-dimensional spinor QCD pertinent for our
canonical analysis both in light-cone and axial gauge.

\section{Quantization of scalar QCD in axial gauge}

We first discuss the general structure of the Hamiltonian in two dimensions for
scalar SU(N) QCD in the axial gauge $A_1=0$. The same discussion for two-dimensional
spinor QCD in both the light-cone and axial gauge is summarized briefly in the Appendix. 
The starting Lagrangian is

\be
\label{1}
{\cal L}=\frac{1}{2}tr F_{01}^2+(D^{\mu}\phi )^{\dagger}D_{\mu}\phi-m^2\phi^{\dagger}\phi
\ee
In terms of the canonical momenta $\pi^{\dagger}=\Pi_{\phi}=(D_0\phi)^{\dagger}$ and $\pi=\Pi_{\phi^{\dagger}}=D_0\phi$, 
the corresponding Hamiltonian reads

\be
\label{2}
H=&& \int dx \bigg(\pi^{\dagger}\pi+|\partial_1 \phi|^2+m^2|\phi|^2\nonumber \\
&&+ig\,{\rm Tr}A_0 (\pi \phi^{\dagger}-\phi \pi^{\dagger})-\frac{1}{2}\,{\rm Tr} (\partial_1 A_0)^2\bigg)
\ee
The equation of motion for $A_0$ is  a constraint equation that can be solved 
in terms of  $\phi, \pi$ , to yield the canonical Hamiltonian

\be
\label{3}
H=&&H_0+H_{int}\nonumber\\
H_0=&&\int dx(\pi^{\dagger}\pi+|\partial_1 \phi|^2+m^2|\phi|^2)\nonumber\\\
H_{\rm int}=&&\frac{g^2}{2}\int dx\left( J^{a}\frac{-1}{\partial_1^2}J^a\right)\nonumber\\
J^a=&&i(\phi^{\dagger}T^a\pi-\pi^{\dagger}T^a\phi)\nonumber\\\
\ee
To proceed, we will use a free-like representation for the field and its conjugate

\be
\label{4}
\phi_\alpha=&&\int\frac{dk}{\sqrt{4\pi E_k}}e^{-ikx}(a_k+b_{-k}^{\dagger})_{\alpha}\nonumber\\
(\pi^{\dagger})_{\alpha}=&&i\int \frac{dk}{\sqrt{4\pi E_k}}e^{ikx}E_k(a_k^{\dagger}-b_{-k})_{\alpha}
\ee
However, instead of the free dispersion law $E_k=\sqrt{k^2+m^2}$, we will use an arbitary 
$E(k)$ that  will be fixed self-consistently below in the planar approximation, with $E_k\rightarrow |k|$
asymptotically.

\subsection{Hamiltonian to order  $1/\sqrt{N_c}$}

The Hamiltonian (\ref{3}) is quartic in $a_k, a^\dagger_k$. We  now choose to bosonize it,
by re-writing it in terms of the quadratic operators

\be
\label{5}
M(k_1,k_2)=&&\frac{1}{\sqrt{N}}\sum_{\alpha}a_{\alpha}(k_1)b_{\alpha}(k_2)\nonumber\\
N(k_1,k_2)=&&\sum_{\alpha}a^{\dagger}_{\alpha}(k_1)a_{\alpha}(k_2)\nonumber\\
\bar N(k_1,k_2)=&&\sum_{\alpha}b^{\dagger}_{\alpha}(k_1)b_{\alpha}(k_2)
\ee
In leading order of $1/\sqrt{N_c}$, 

\be
\label{6}
N(k,p)=&&\int dq M^{\dagger}(k,q)M(p,q)\nonumber\\
\bar N(k,p)=&&\int dq M^{\dagger}(q,k)M(q,p)
\ee
Using (\ref{5}-\ref{6}) and the identity $\sum_a (T^a)_{ij} (T^a)_{kl}=\delta_{il}\delta_{kj}-\frac{1}{N}\delta_{ij}\delta_{kl}$,
the Hamiltonian (\ref{3}) now reads to order $1/\sqrt{N_c}$ as

\be
\label{7}
&&H=H_2+H_4\nonumber\\
&&H_2=\int dk (N(k)+\bar N(k))\Pi^{+}(k)\nonumber \\ 
&&+\sqrt{N_c}\int dk (M(k)+M^{\dagger}(k))\Pi^{-}(k)\nonumber\\
&&H_4=\lambda \int \frac{dk_1dk_2dk_3 dk_4}{16\pi}\frac{\delta(k_1+k_2+k_3+k_4)}{(k_1+k_2)^2}\nonumber \\ 
&&\times \bigg(-2f_+(k_1,k_2)f_+(k_3,k_4)M^{\dagger}(k_1,k_4)M(-k_2,-k_3)\nonumber \\
&&+f_-(k_1,k_2)f_-(k_3,k_4)M^{\dagger}(k_1,k_4)M^{\dagger}(k_3,k_2)\nonumber \\ 
&&+f_-(k_1,k_2)f_-(k_3,k_4)M(k_1,k_4)M(k_3,k_2))\nonumber\\
&&+{\cal O}\left(\frac{1}{\sqrt{N_c}}\right)\bigg)
\ee
Here $\lambda=g^2N_c$  is the standard t$^\prime$ Hooft coupling. We have made use use of the notation 
$M(k)=M(k,-k)$, $N(k)=N(k,k)$, and

\be
\label{8}
&&f_{\pm}(k_1,k_2)=\sqrt{\frac{E_2}{E_1}}\pm \sqrt{\frac{E_1}{E_2}}\nonumber\\
&&\Pi^{\pm}=\frac{1}{2}\bigg(\frac{k^2+m^2}{E_k}\pm E_k\bigg)+\lambda \int\frac{dk_1}{8\pi}\frac{\frac{E_{k_1}}{E_k}\pm \frac{E_{k}}{E_{k_1}}}{(k+k_1)^2}
\ee
For a consistent expansion in $1/N_c$, we can eliminate the $\sqrt{N_c}$ term in (\ref{7}) by setting 
$\Pi^{-}(k)=0$. The result is a gap equation for  $E(k)$

\be
\label{9}
\frac{k^2+m^2}{E_k}-E_k+\frac{\lambda}{4\pi}\int dk_1\bigg(\frac{E_{k_1}}{E_{k}}-\frac{E_{k}}{E_{k_1}}\bigg)\frac{1}{|k+k_1|^2}=0\nonumber \\ 
\ee
The leading order Hamiltonian simplifies to

\be
\label{10}
&&H=\int dpdqM^{\dagger}(p,q)M(p,q)(\Pi^{+}(p)+\Pi^{+}(q))\nonumber \\
&&+\lambda \int \frac{dk_1dk_2dk_3 dk_4}{16\pi}\frac{\delta(k_1+k_2+k_3+k_4)}{(k_1+k_2)^2}\nonumber \\
&& \times \bigg(-2f_+(k_1,k_2)f_+(k_3,k_4)M^{\dagger}(k_1,k_4)M(-k_2,-k_3)\nonumber \\
&&+f_-(k_1,k_2)f_-(k_3,k_4)M^{\dagger}(k_1,k_4)M^{\dagger}(k_3,k_2)\nonumber \\
&&+ f_-(k_1,k_2)f_-(k_3,k_4)M(k_1,k_4)M(k_3,k_2)\bigg)
\ee

\subsection{Renormalization}

The integral in the gap equation (\ref{9}) and subsequently  the Hamiltonian contains a divergence and requires
 regularization.   For that we regularize  $\frac{1}{(k+k_1)^2}$ using the standard principal value (PV) prescription

\be
\label{R1}
\int dx \frac{f(x)}{(x-y)^2}\rightarrow {\rm PV}\int dx \frac{f(x)-f(y)}{(x-y)^2}+\frac{2f(y)}{\epsilon}
\ee
It is readily seen that $\Pi^-$ is finite but $\Pi^+$ diverges as

\be
\label{R2}
\Pi^{+}=\Pi^{+}_r+\frac{\lambda}{2\pi \epsilon}
\ee
with $\Pi_r$ finite.  We have checked that for physical states (on mass shell) the $\epsilon$-contributions
cancel out (see below).

The solution to the gap equation (\ref{9}) that asymptotes $E_k\rightarrow |k|$ still suffers from a logarithmic divergence
even after the PV prescription, namely

\be
\label{R3}
\frac{\lambda}{8\pi E_k} \int dk_1 \frac{E_{k_1}}{k_1^2}
\ee
This is actually related to the mass divergence for the scalar one-loop self energy,  and renormalizes  the scalar mass

\be
\label{R4}
m_r^2=m^2+\frac{\lambda}{4\pi } \int dk_1 \frac{E_{k_1}}{k_1^2}
\ee
From here on, we will refer to $\Pi^+$ as the renormalized momentum operator, and $m$ as the renormalized mass,
 and omit the r-label for convenience.  With this in  mind, the renormalized gap equation (\ref{9}) now reads

\be
\label{R5}
&&\frac{k^2+m^2}{E_k}-E_k+\nonumber \\ 
&&\frac{\lambda}{4\pi}\int dk_1\bigg(\bigg(\frac{E_{k_1}}{E_{k}}-\frac{E_{k}}{E_{k_1}}\bigg)\frac{{\rm PV}}{|k+k_1|^2}- \frac{E_{k_1}}{E_k}\frac 1{k_1^2}\bigg)=0\nonumber \\
\ee

\section{Wave-function and light cone limit}

To construct the light cone wave-function of the scalar quarks, we define

\be
\label{11}
&&S_{\pm}(p,k,P)=f_{\pm}(p-P,k-P)f_{\pm}(p,k)\nonumber\\
&&A=2S_{+}(p,k,P)M^{\dagger}(p-P,p)M(k-P,k)\nonumber\\
&&B=S_{-}(p,k,P)(M^{\dagger}(p,p-P)M^{\dagger}(k-P,k)+c.c)\nonumber\\
\ee
and use them  to re-write (\ref{10}) in the form

\be
\label{12}
H=&&\int dpdqM^{\dagger}(p,q)M(p,q)\,(\Pi^{+}(p)+\Pi^{+}(q))\nonumber \\
&&-\frac{\lambda}{16\pi}\int dP \int {dkdp}\,\frac {A+B}{(p-k)^2}
\ee
The bi-local operator $M(p,q)$ can be decomposed in modes

\be
\label{13}
&&M(p-P,p)=\nonumber \\ 
&&\frac{1}{\sqrt{|P|}}\sum_{n}\bigg(m_n(P)\phi_{n}^{+}(q,P)-m_{n}^{\dagger}(-P)\phi_n^{-}(q-P,-P)\bigg)\nonumber\\
\ee
where the first contribution refers to the light cone wavefunction describing a pair of scalar quarks
moving forward in the light front, while the second contribution refers to a pair moving backward in the
front form. The pair is characterized by a relative momentum $p$ and a center of mass momentum $P$. 
Here $m_n, m_n^\dagger$ are canonical bosonic annihilation and creation operators. The
equation of motion follows by commutation

\be
\label{14}
&&(\Pi^{+}(p)+\Pi^{+}(P-p)\mp P_n^{0})\phi_{n}^{\pm}(p,P)=\nonumber \\ 
&&\frac{\lambda}{8\pi}\int\frac{dk}{(p-k)^2}\nonumber\\
&&\times (S_{+}(p,k,P)\phi_n^{\pm}(k,P)-S_{-}(p,k,P)\phi_n^{\mp}(k,P))\nonumber\\
\ee

We can check that the $\epsilon$-dependent divergences noted in the momentum operator
cancel out. Indeed,  using (\ref{R2}) the LHS in (\ref{14}) produces $\frac{\lambda}{\pi \epsilon} \phi^{\pm}$, while
the RHS in (\ref{14}) produces $\frac{\lambda}{4\pi \epsilon}S^{+}(k,k)\phi^{\pm}=\frac{\lambda}{\pi \epsilon}\phi^{\pm}$,
both of which cancel out. This checks  the consistency of the renormalization procedure for scalar QCD.
No such renormalization is needed for spinor QCD. 

In the large momentum limit $P$ the equation simplifies. For that we set 
$p=xP$,  $k=yP$, and take $P\rightarrow \infty$ on both sides of (\ref{14}).
In this limit the backward wavefunction vanishes $\phi_{-}\rightarrow 0$.
Since

\be
\label{15}
&&\Pi^{+}(Px)+\Pi^{+}((1-x)P)-\sqrt{P^2+M^2}\nonumber \\
&&=\frac{1}{2P}\bigg(\frac{m^2}{x}+\frac{m^2}{1-x}-M_n^2\bigg)+{\cal O}\bigg(\frac{1}{P^2}\bigg)
\ee
and

\be
\label{16}
S_{+}(xP,yP,P)=\frac{(2-x-y)(x+y)}{\sqrt{x(1-x)y(1-y)}}
\ee
the equation of motion (\ref{14}) involves only the forward wavefunction in the form

\be
\label{17}
&&\bigg(\frac{m^2}{x}+\frac{m^2}{1-x}-M_n^2\bigg)\phi_{n}(x)=\nonumber \\ 
&&\frac{\lambda}{4\pi}{\rm PV}\int \frac{dy}{(x-y)^2}\frac{(2-x-y)(x+y)}{\sqrt{x(1-x)y(1-y)}}\phi_n(y)
\ee
where we have defined $\phi_{n}^{+}(xP,P)=\phi_{n}(x)$,
and PV refers to the principal value of the integral. (\ref{17}) was obtained initially 
in~\cite{TAM} using different arguments.

\section{quasi-parton distribution function}

The light cone distribution for scalar quarks is just 
$|\phi_n(x)|^2$  in leading order in $1/N_c$. We now show that to the same order,
the light cone distribution function and the quasi-distribution function as defined
in~\cite{JI1} are in agreement without further normalization. For that, we define the
quasi-distribution function

\be
\label{18}
\tilde q(x,P)=&&+i\int \frac{dz}{4\pi}e^{iPxz}\left<P|(\partial_1 \phi(z))^{\dagger}W[z,0]\phi(0)|P\right>\nonumber \\ 
&&-i\int \frac{dz}{4\pi}e^{iPxz}\left<P|(\phi(z))^{\dagger}W[z,0]\partial_1\phi(0)|P\right>\nonumber\\
\ee
where $\left|P\right>$ refers to the meson state.  In the axial gauge, the Wilson line $W[z,0]=1$.
Using the mode decomposition (\ref{4}) and the relations (\ref{6}) we obtain for the quasi-distribution

\be
\label{19}
\tilde q(x,P)=&&\frac{E_n(P)}{P}\frac{xP}{E(xP)}\nonumber \\ 
&&\times \bigg(|\phi^{+}_n(xP,P)|^2+|\phi^{+}_n(-xP,P)|^2\nonumber \\
&& +|\phi^{-}_n((xP,P)|^2+|\phi^{-}_n(-xP,P)|^2\bigg)
\ee
For $P\rightarrow \infty$, we have $E_{n}=P$ and $xP=E(xP)$ and all $\phi_{-}$ vanish.
The quasi-parton distribution function reduces identically to the parton distribution function 
$|\phi_{n}(x)|^2$.

For finite $P$, (\ref{19}) shows that the backward moving pair in $\phi^-$ contributes. To assess
this quantitatively, 
we now expand in $\frac 1P$ the contributions $\phi^\pm$ in (\ref{19}). For that, we go back to 
(\ref{14})  and expand in $\frac 1P$, namely

\be
\label{19X}
\Pi^{+}=&&|P|+\frac{m^2}{2|P|}+\frac{\beta_1}{2|P|^3}+{\cal O}\left(\frac 1{|P|^4}\right)\nonumber\\
E(P)=&&|P|+\frac{\beta_2}{|P|}+{\cal O}\left(\frac 1{|P|^2}\right)
\ee
The coefficients $\beta_{1}$ is fixed through a straightforward Taylor expansion of $\Pi^+$, 
while $\beta_2$ is  fixed by the gap equation. Their explicit form is not needed for the
general arguments to follow. With this in mind, the leading correction to $\phi^-$ is

\be
\label{19XX}
\phi^-_1=P^2\phi_n^{-}(x)=\frac{\lambda}{24\pi \sqrt{x(1-x)}}\int_0^1 \frac{dy\phi_n(y)}{\sqrt{y(1-y)}}
\ee
and the subleading correction for $\phi^{+}=\phi(x)+\frac{1}{P^2}\phi^{+}_1(x)$ formally solves

\be
\label{19XXX}
\tilde \phi\equiv (K_0 -H_0)\phi_1^+=-K_1\phi+H_1\phi-H^{-}_0\phi_1^-
\ee
Here we have defined

\be
\label{19Z}
&&K_0(x)=\frac{m^2}{x}+\frac{m^2}{\bar x}-M_n^2\nonumber\\
&&H_0(x,y)=\frac{\lambda}{4\pi}\frac{(\bar x+\bar y)(x+y)}{\sqrt{xy\bar x\bar y}}\frac 1{(x-y)^2}\nonumber\\
&&K_1(x)=\frac{\beta_1}{x^3}+\frac{\beta_1}{\bar x^3}\nonumber\\
&&H_0^-(x,y)=-\frac{\lambda}{4\pi\sqrt{xy\bar x \bar y}}\nonumber\\
&&H_1(x,y)=\frac 1{(x-y)^2\sqrt{x\bar xy\bar y}}\nonumber \\ 
&&+\beta_2(x^2-y^2)\bigg(\frac{1}{\bar y^2}-\frac{1}{\bar x^2}\bigg)+\beta_2(\bar x^2-\bar y^2)\bigg(\frac{1}{y^2}-\frac{1}{x^2}\bigg)\nonumber\\
\ee
with $\bar x=1-x$ and $\bar y =1-y$.
In general , this equation is solved in the same Hilbert space that defines $K_0-H_0$, 
if we note that $K_0-H_0$ is hermitian in the space defined with the measure
$\int \phi^{\dagger}\phi$ where the set of $\phi_n$ forms a complete basis set. The formal solution to
(\ref{19Z}) is

\be
\phi_1^{+}(x)=\sum_{m \ne n}\frac{\phi_m(x)\int_0^1 dy \phi_m^{\dagger}(y) \tilde \phi(y)}{M_m^2-M_n^2}
\ee

The $\frac 1P$ expansion now clearly shows that the the rate at which the quasi-distribution (\ref{19}) 
approaches the asymptotic light-cone distribution $|\phi_n(x)|^2$ is smooth for all $x\neq 0, 1$. It is singular for $x=0,1$
through the contribution of the backward moving pair $\phi^-$ in (\ref{19XX}). So the large $P$ limit should
be taken before the $x\rightarrow 0,1$ limits at the edges.

\section{Algebraic structure}

The algebraic framework we have developed allows us to go beyond the leading order in $1/N_c$,
and therefore check the proposal in~\cite{JI1} beyond the leading order we have so far established. 
For that, we note that the bi-local operators (\ref{5}) obey a closed algebra

\be
\label{20}
&&[M_{12},M^{\dagger}_{34}]\nonumber \\ 
&&=\delta_{13}\delta_{24}+ \frac{s}{N_c}(\delta_{13}\bar N_{42}+\delta_{42}N_{31})\nonumber\\
&&[M_{12},N_{34}]=\delta_{13}M_{42}\nonumber\\
&&[M_{12},\bar N_{34}]=\delta_{23}M_{14}\nonumber\\
&&[M_{12},M_{34}]=[N_{12},\bar N_{34}]=0\nonumber\\
&&[N_{12},N_{34}]=\delta_{23}N_{14}-\delta_{14}N_{32}
\ee
with $N^{\dagger}_{12}=N_{21}$. The sign assignment for the bosonization of scalar QCD is $s=+1$
as all underlying operators are bosonic.

A solution to this algebraically closed set can be found by  organizing the bi-local operator in $1/N_c$,

\be
\label{21}
&&M=M^{0}+\frac{1}{N_c}M^1+{\cal O}\bigg(\frac{1}{N_c^2}\bigg)\nonumber\\
&&N=N^{0}+\frac{1}{N_c}N^1+{\cal O}\bigg(\frac{1}{N_c^2}\bigg)
\ee
where $M^0$ satisfies the commutation relation 

\be
\label{22}
[M^0(k_1,k_2),M^{0\dagger}(k_3,k_4)]=\delta(k_1-k_3)\delta(k_2-k_4)
\ee
 in the infinit $N_c$ limit. In terms of (\ref{21}-\ref{22}) the solution to (\ref{20}) 
 can be found by inspection in leading and next to leading order

\be
\label{23}
N^{0}_{12}=&&\int d3 M^{0\dagger}_{13}M^0_{23}\nonumber\\ 
\bar N^{0}_{12}=&&\int d3 M^{0\dagger}_{31}M^0_{32}\nonumber\\
M^{1}_{12}=&&\mp \frac{1}{2}\int d3d4 M^{0\dagger}_{34}M^{0}_{14}M^{0}_{32}\nonumber\\
N^{1}=&&0
\ee
It is important to note that the expantion of the $N$'s starts at the second order! From now on to avoid cluttering,
we omit the 0 for the large $N_c$ asymptotic operator.

When the operators in (\ref{23}) are inserted back into the Hamiltonian, 
we obtain a complete expression for the first three terms  of the $1/N_c$ expanded Hamiltoinian in terms 
of the large $N_c$ asymptotic operators that define the Hilbert space. Specifically, to order $\frac{1}{N_c^2}$ we have

\be
\label{24}
&&H=K_{MM}M^{\dagger}M \nonumber \\ 
&&+\frac{1}{N_c}K_{MM}(M^{\dagger 1}M+M^{\dagger}M^1)+\frac{1}{N_c^2}K_{MM}M^{1\dagger}M^{1}\nonumber \\
&&+\frac{K_{NM}}{\sqrt{N_c}}NM+\frac{K_{NM}}{N_c\sqrt{N_c}}NM^1+\frac{K_{NN}}{N_c}NN
\ee
Thus, up to order ${1}/{N_c^2}$ we encounter six $M$ interactions, but up to oder ${1}/{N_c\sqrt{N_c}}$ we are still dealing with more tractable quartic and qubic terms. Our algebraic treatment differs notablly from the one presented in \cite{BARBON} in that in ours
the algebra is corrected which is required for a consistent expansion. The resulting effective hadronic Hamiltonian is different.

\section{Correction to the PDF  in spinor QCD}

In so far our discussion has concentrated on two-dimensional scalar QCD where we have established that
the quasi-parton distribution function reduces to the parton distribution function in leading order in $1/N_c$.
We have checked that this is also the case for two-dimensional spinor QCD, in agreement with a recent
study~\cite{RECENT}.
In the Appendix we have briefly summarized the key changes from scalar to spinor in the light cone and 
axial gauge.

Since in the spinor version,  the underlying fields are fermionic and not bosonic, the algebraic structure
(\ref{20}) differs from scalar to spinor QCD  only in the sign switch $s=+1\rightarrow -1$, with exactly the same bosonized
Hamiltonian (\ref{24}). Also, 
to avoid unecessary long formula we will only discuss the $1/N_c$ corrections to the parton distribution
function in two-dimensional spinor instead of scalar QCD. The  arguments for both models are similar,
but the formula for scalar QCD  are laboriously long as we have checked, with exactly the same conclusion.

Using the definitions for spinor QCD in the
Appendix, we use for the bi-local mesonic operator $M$ in the light cone gauge the decomposition

\be
\label{25}
M(xP,(1-x)P)=\frac{1}{\sqrt{P}}\sum_{n}m_{n}(P)\phi_n(x)
\ee
which satisfies (\ref{20}) with $s=-1$.  To order $1/N_c$, the Hamiltonian for two-dimensional spinor QCD 
is the same as in (\ref{24}),  which after inserting (\ref{25}) yields the first two leading contributions to the interaction
of the form

\be
\label{26}
&&\frac{\lambda}{4\pi\sqrt{N_c}}\int \frac{dPdP_1}{P^{\frac{3}{2}}}\nonumber\\
&&\times\bigg( m_{i}^{\dagger}(P_1)m_{j}^{\dagger}(P-P_1)m_{k}(P)f_{ijk}(\frac{P_1}{P})+{\rm c.c}\bigg)\nonumber\\
&&+\frac{1}{N_c}m^{\dagger}m^{\dagger}mm  
\ee
The quartic contribution in (\ref{26}) is only shown schematically. It is of order $1/N_c$, and apparently
relevant for the $1/N_c$ correction to the parton distribution function. However, by simple inspection it
gives zero contribution when acting on a free and leading meson contribution to the state, i.e.

\be
\label{27}
\bigg( \frac 1{N_c}\,m^{\dagger}m^{\dagger}mm \bigg)\,m^\dagger \left|0\right>=0
\ee 
It will be dropped. Therefore the leading correction to the parton distribution function is given by

\be
\label{28}
&&\sum_{kl}\int \frac{dk dq}{2\pi}\phi _{k}\left(\frac{xP}{xP+q}\right)\phi_{l}\left(\frac{k}{k+q}\right)\nonumber \\ 
&&\times ^{1}\left<P_i\right|\bigg(\frac{m_{k}^{\dagger}(xP+q)m_{l}(k+q)}{\sqrt{(xP+q)(k+q)}}\bigg)\left|P_i\right>^{1}
\ee
Here $\left|P\right>^{1}$ is the first order perturbation of the meson state $m_{i}^{\dagger}(P)\left|0\right>$, 
which by standard perturbation theory reads

\be
\label{29}
&&\left|P\right>^{1}=\frac{\lambda}{2 \sqrt{2\pi N_c}}\int dP_1 \nonumber\\
&&\times \sum_{kl}
\bigg(\frac{f_{kli}(\frac{P_1}{P})m_{k}^{\dagger}(P_1)m_{l}^{\dagger}(P-P_1)}{\frac{m_k^2}{x}+\frac{m_l^2}{1-x}-m_i^2}\bigg)
\left|0\right>
\ee
Inserting (\ref{29}) into (\ref{28}) and carrying out the contractions yield 

\be
\label{30}
\delta q_i(x)=\frac{\lambda^2}{\pi^2 N_c}\int_0^{1-x}dy\sum_{kk^{\prime}l}\frac{F_{kli}(x,y)F_{k^{\prime }li}(x,y)}{x+y}
\ee
as a correction to the leading parton distribution function $q_i(x)=|\phi_i(x)|^2$, with 

\be
\label{31}
F_{kli}(x,y)=\frac{f_{kli}(x+y)\phi_k(\frac{x}{x+y})}{\frac{m_k^2}{x+y}+\frac{m_l^2}{1-x-y}-m_i^2}
\ee
and

\be
\label{32}
&&\frac{f_{ijk}(x)}{\sqrt{x(1-x)}}=\nonumber \\ 
&&\int dx_1 dx_2\frac{\phi_i(x_1)\phi_j(x_2)\phi_k(x+x_2-xx_2)}{(xx_1+xx_2-x-x_2)^2} \nonumber \\
&&-\int dx_1 dx_2\frac{\phi_i(x_1)\phi_j(x_2)\phi_k(x_2-xx_2)}{(xx_1+xx_2-x-x_2)^2} 
\ee

\section{Correction to the quasi-PDF in spinor QCD}

In this section we derive the $1/N_c$ correction to the quasi-parton distribution function for two-dimensional
spinor QCD and show that it is in agreement with the $1/N_c$ correction to the parton distribution we just
established in the large momentum limit. For  that, we switch to the description of two-dimensional spinor
QCD in the axial gauge using the changes in the Appendix.

In the axial gauge, the Hamiltonian is writtent in
terms of  $m_n(P)$ and $\phi_{\pm}$. The structure of the Hamiltonian is still of the form (\ref{24}).
We now note that the contributions to the first order shift of the state $\left|P\right>^1$ of the form
$m^{\dagger}m^{\dagger}m^{\dagger}$ always carries $\phi^{-}$. In the large momentum limit these
terms drop out as we have shown earlier, so they will be ignored. The only surviving
terms in the Hamiltonian at large momentum are also of the form $m^{\dagger}m^{\dagger}m+{\rm c.c}$.  

With the above in mind and to be more specific, the parts of the Hamiltonian (\ref{24}) that will 
contribute to the quasi-parton distribution function in leading order in perturbation theory are
of the form

\be
\label{33}
&&H_{1}=\frac{1}{\sqrt{N_c}}\sum_{123}f_{123}m_1^{\dagger}m_2^{\dagger}m_3\nonumber\\
&&H_{2}=\frac{1}{N_c}\sum_{1234}f_{1234}m_1^{\dagger}m_2^{\dagger}m_3^{\dagger}m_4
\ee
The ensuing shifts caused by (\ref{33}) on the mesonic state  to first order in $\frac{1}{N_c}$
are respectively of the form

\be
\label{34}
&&\left|i\right>^{1}=\frac{1}{\sqrt{N_c}}\sum_{12}\left|12\right>\alpha_{12i}\nonumber\\
&&\left|i\right>^{2}=\frac{1}{N_c}\sum_{123}\left|123\right>\alpha_{123i}
\ee
with

\be
\label{35}
&&\alpha_{12i}=\frac{f_{12i}}{E_1+E_2-E_i}\nonumber\\
&&\alpha_{123i}=\frac{f_{123i}}{E_1+E_2+E_3-E_i}\nonumber \\ 
&&+\sum_4 \frac{f_{123}f_{34i}}{(E_1+E_2+E_3-E_i)(E_3+E_4-E_i)}
\ee
and the coefficients $f_{ijk}$ and $f_{ijkl}$ are 

\be
\label{36}
&&f_{ijk}(P_1,P_2,P_3)= \frac{\lambda}{4\pi}\int dk_1dk_2dk_3dk_4dq \nonumber \\
&&\times\delta(k_1+k_2+k_3-k_4) \nonumber \\
&&\times \delta(k_1+k_2-P_1)\delta(k_3+q-P_2)\delta(k_4+q-P_3)\nonumber \\
&&\times\bigg( \frac{\phi^{+}_i(k_1,P_1)\phi^{+}_j(k_3,P_2)\phi^{+}_k(k_4,P_3)S(k_1,k_2,k_3,k_4)}{(k_1-k_4)^2}\nonumber \\
&& -\frac{\phi^{+}_i(k_1,P_1)\phi^{+}_j(q,P_2)\phi^{+}_k(q,P_3)S(k_2,k_1,k_3,k_4)}{(k_1+k_3)^2}\bigg)\nonumber \\ 
&&+ f_{ijk}^{-}
\ee
where  we have set 

\be
\label{37}
S(k_1,k_2,k_3,k_4)=\cos \frac{\theta(k_1)-\theta(k_4)}{2}\sin \frac{\theta(k_2)+\theta(k_3)}{2} \nonumber\\
\ee
The last contribution $f_{ijk}^-$  involves at least one $\phi^-$ and therefore drops out in the 
large momentum limit, so it will not be quoted.

All  contributions of the form $f_{ijkl}$ 
involve at least one $\phi^{-}$ and also drop out in the large momentum limit. 
More specifically, in the large momentum limit, we set $P_i=P\rightarrow +\infty$, and we change our variables to $P_1=xP$, $P_2=yP$, and  $P_3=zP$, then any term which contains $\phi^{-}(x_1P,x_2P)$ vanishes in this limit,  an example is  the $f_{1234}$ term.

The parton fractions are constrained kinematically. For instance, the energy denominator

\be
\frac{1}{E_{xP}+E_{yP}+E_{z}-E_p}\nonumber
\ee
implies  $0<x,y,z<1$ in leading order in $1/P$, otherwise the contribution  is subleading. 
In this case, the only term in $H^{1}$ which contains only $\phi^{+}$ (first contribution in (\ref{24}))
will reduce to the light cone gauge term if one identifies
the creation operators in both cases using

\be
\label{38}
&&\phi^{+}_n(xP,P)\rightarrow \phi_n(x)\nonumber\\
&&\frac{1}{E_{xP}+E_{(1-x)P}-E_P}\rightarrow \frac{2P}{\frac{m_1^2}{x}+\frac{m_2^2}{1-x}-m_i^2}
\ee
More specifically, the first order correction to the quasi-parton distribution function is proportional to

\be
\label{39}
&&\left<P\right|\int dp dq \sin \left[\frac{\theta(xP)+\theta(p)}{2}\right]M^{\dagger}(xP,q)M(p,q)\left|P\right>+\nonumber\\ 
&&\left<P\right|\int dp dq \sin \left[\frac{\theta(xP)+\theta(p)}{2}\right]M^{\dagger}(q,-p)M(q,-xP)\left|P\right>\nonumber\\
\ee
with $\left|P\right>$ corrected to first order.  There are two type of contributions in (\ref{39})
as we now discuss.

First, the $m^{\dagger}m$ term. For this only the $\left|i\right>^{1}$ in the shift of the state 
contributes, and the specific contribution  with only $\phi^{+}$ is

\be
\label{40}
&&2\sin \frac{\theta(xP)}{2}\sum_{kk^{\prime}l}\frac{\alpha_{kli}\alpha_{k^{\prime }li}}{|k|} \phi^{+}_k(xP,p_k)\phi^{+}_{k^{\prime}}(xP,p_k)\nonumber \\ 
&&+2\sin \frac{\theta(xP)}{2}\sum_{kk^{\prime}l}\frac{\alpha_{kli}\alpha_{k^{\prime }li}}{|k|} \nonumber\\
&&\times \phi^{+}_k(xP+p_k,p_k)\phi^{+}_{k^{\prime}}(xP+p_k,p_k)\nonumber \\
\ee
In the large momentum limit, we have $p_k=yP$,  and $p_l=(1-y)P$ as discussed above.
The first term is non-zero if $0<x<y$, and the second term is always zero for $0<x<1$ since  $(x+y)>y$. 
Thus by shifting $y\rightarrow y+x$ with $0<y<1-x$, and taking care of factors of $P$,  this contribution matches
the correction to the parton distribution function in the light cone gauge (\ref{30}).

Second, the $mm+m^{\dagger}m^{\dagger}$ term comes with at least one $\phi^{-}$, and is always zero in the large $P$ limit as discussed above. It follows, that the order $1/N_c$ contribution to the quasi-parton distribution matches the parton distribution
in the large momentum limit without renormalization in two-dimensional spinor QCD. We have explicitly checked that the same holds 
for two-dimensional scalar QCD.

\section{Conclusions}

Using a bosonized form of two-dimensional scalar and spinor QCD, we have analyzed the quasi-parton 
distribution of a meson state. In the infinite momentum limit, the quasi-distribution matches the parton 
distribution on the light cone both in leading and sub-leading order without further renormalization, 
but the limit is subtle at the parton fractions $x=0,1$. 
This provides a non-perturbative check on the proposal put forth by one of us~\cite{JI1} for extracting the QCD light
cone partonic distributions from their quasi-distribution counterparts using pertinent  equal-time Euclidean correlators
through suitable matching at large momentum.

\section{Acknowledgements}

This work was supported by the U.S. Department of Energy under Contract No.
DE-FG02-93ER-40762 for XJ and DE-FG-88ER40388 for YL and IZ.

\section{Appendix :  two-dimensional spinor QCD in the light-cone and axial gauge}

Here and for convenience we briefly summarize some of the changes needed to
recover spinor QCD from scalar QCD as developed in the main text. Both in the light cone and axial gauge
 the mesonic operators $M$ and $N$ are defined as in section ${\bf V}$ with $s=-1$. The fermionic fields
 in terms of creation-annihilation operators are defined as

\be
\label{41}
&&\psi=\int_{0}^{\infty} \frac{dp^{+}}{2\pi}(a(p^+)e^{-ip^+x^-}+b^{\dagger}(p^+)e^{-ip^+x^-})\nonumber\\
&&\psi=\int\frac{dp}{2\pi}e^{ipx}(a(p)u(p)+b^{\dagger}(-p)v(-p))
\ee
in the light cone and axial gauge respectively, with 

\be
\label{42}
&&u(p)=e^{-\frac{1}{2}\theta(p)\gamma^1}(1,0)^T\nonumber\\
&&v(-p)=e^{-\frac{1}{2}\theta(p)\gamma^1}(0,1)^T
\ee
The mode decomposition in the light cone gauge is given in (\ref{25}), and in the aixal gauge as

\be
\label{43}
&&M(k_1,P-k_1)=\frac{1}{\sqrt{|P|}}\nonumber \\ 
&&\times \sum_n \phi^+_n(k_1,P)m_n(P)-\phi^-_n(-k_2,-P)m^{\dagger}_n(-P)
\ee
The bosonized Hamiltonian is still of the form (\ref{24}), 
with the relevant $M^{\dagger}MM$ term given in the main text.



 \vfil

\end{document}